\documentclass[a4paper,11pt]{article}
\usepackage{pos}
\usepackage{slashed}
\usepackage{subcaption}
\usepackage{bbm}

\def\tb{{\bar{t}}}
\def\eps{\epsilon}

\def\mren{\mathrm{mren}}

\def\cO{\mathcal{O}}

\def\la{\langle}
\def\ra{\rangle}
\def\spA#1#2{\la#1#2\ra}
\def\spB#1#2{[#1#2]}

\def\spAA#1#2#3{\la#1|#2|#3\ra}

\def\fl#1{#1^\flat}

\newcommand{\qb}{{\bar{q}}}

\DeclareMathOperator{\Tr}{Tr}
\title{Analytic representations of two-loop scattering amplitudes with internal masses}
\author*{Ekta Chaubey}
\affiliation{Physics Department, University of Turin and INFN Turin\\
  Via Pietro Giuria 1, I-10125, Turin, Italy}

\emailAdd{ekta@to.infn.it}

\abstract{We highlight the latest developments in computing higher-order
scattering amplitudes with massive internal propagators. The contributing Feynman
integrals often lead to special classes of functions, for example,
functions associated with elliptic curves. With the presence of more
scales in the amplitudes, it becomes imperative to have a better
understanding of the contributing Feynman integrals and using current cutting-edge technologies to tackle the growth in analytic and algebraic complexities. In particular, we start with discussing two-loop scattering amplitudes for top-quark pair production and conclude with motivating important steps towards obtaining next-to-next-to leading-order corrections for five-point processes.
 }

\FullConference{%
  Loops and Legs in Quantum Field Theory - LL2022,\\
  25-30 April, 2022\\
  Ettal, Germany
}

\begin{document}
\maketitle

Precision measurements for top-quark pair production are one of the main priorities for the Large Hadron Collider (LHC). Measurements of top-quark pair (and single top quark) cross sections can test the Standard Model (SM) and probe new physics. The Higgs potential is very sensitive to the top quark mass. Top quark processes also constitute an important background of many searches and measurements. Therefore it is imperative to carry out higher-order precision calculations for processes involving top quarks.

To compute  precise higher-order contributions for phenomenological applications, we need to calculate high-multiplicity scattering amplitudes. For massless scattering amplitudes, there have been impressive results coming in from the past few years, see for example \cite{Badger:2022ncb,Chawdhry:2021mkw,Sotnikov:2022mzg}. To contribute further to the world of precision calculations, we need to include corrections due to massive particles. To obtain next-to-next-to (NNLO) corrections for $pp \rightarrow t\bar{t}$, we need 2-loop 2$\rightarrow$ 2, 1-loop 2$\rightarrow$3 and tree-level 2$\rightarrow$4 partonic contributions. For NNLO corrections to $pp \rightarrow t\bar{t} j$, 2-loop 2$\rightarrow$3, 1-loop 2$ \rightarrow$4 and tree-level 2$\rightarrow$5 contributions are needed whereas for $\text{N}^3$LO corrections to $pp \rightarrow t \bar{t}$ we need all 3-loop 2$\rightarrow$2, 2-loop 2$\rightarrow$3 , 1-loop 2$\rightarrow$4 and tree-level 2$\rightarrow$6 contributions. For the inclusive top-quark pair production, fully differential NNLO predictions using numerical approaches are available for precise comparisons with data \cite{Barnreuther:2012wtj,Czakon:2012zr,Czakon:2012pz,Czakon:2013goa,Barnreuther:2013qvf,Chen:2017jvi}. One of the main reasons for using a numerical approach for these computations is that with the inclusion of masses the analytic structure of the amplitudes becomes quite challenging to tackle as often integrals evaluating to a more complicated class of functions (for example functions associated with elliptic curves) start to appear \cite{Adams:2018bsn,Broedel:2019kmn}. Proper handling of these structures requires more mathematical understanding. There has been substantial progress over the last few years in the world of elliptics \& beyond \cite{Bourjaily:2022bwx,Muller:2022gec}.

In these proceedings, we discuss the NNLO analytic contributions for two-loop helicity amplitudes for top-quark pair production in the leading color approximation, as well as an important NNLO ingredient for analytic corrections for $t\bar{t}j$ production. We elaborate on the complexities arising due to integrals and the choice of special functions, as well those from analytic coefficients and high degrees of rational functions.

\section{Methodology}
\label{sec:method}
The computation of higher order perturbative corrections contains a large number of steps with different bottlenecks. We make use of finite field arithmetic for an efficient solution to the large system of integration-by-parts identities. Apart from being efficient for massless amplitudes with many external scales, we have shown that the techniques that we use are equally compatible with multi-leg scattering amplitudes with massive internal propagators. In both of the processes discussed here, we make use of rational parametrization construction of the on-shell kinematics via momentum twistors. We also obtain helicity amplitudes by projecting them onto an independent basis of spinor structures that account for the freedom in the top quark spin states. With this method, we can define a set of on-shell, gauge invariant sub-amplitudes which can be computed using on-shell top quark kinematics. These can be used to describe the full set of spin-correlated narrow width decays.

We start by reviewing one way to incorporate massive spinors into spinor helicity techniques. The first step is to introduce an arbitrary direction $n$ used to define a massless projection of the massive fermion momentum
\begin{equation}
  p^{\flat,\mu} = p^\mu - \frac{m^2}{2p.n} n^\mu.
  \label{eq:pflatdef}
\end{equation} 
Here $p^\mu$ is the momentum of an on-shell massive fermion with $p^2 =m^2$, where both $p^{\flat}$ and $n$ are both massless. We then construct $u$ and $v$ spinors for massive fermions using Weyl spinors for massless momenta,
\begin{align}
  u_+(p,m) &= \frac{(\slashed{p}+m)|n\ra}{\spA{\fl{p}}{n}}, &
  u_-(p,m) &= \frac{(\slashed{p}+m)|n]}{\spB{\fl{p}}{n}},
  \label{eq:massivespinors1}  \\
  v_-(p,m) &= \frac{(\slashed{p}-m)|n\ra}{\spA{\fl{p}}{n}}, &
  v_+(p,m) &= \frac{(\slashed{p}-m)|n]}{\spB{\fl{p}}{n}}.
  \label{eq:massivespinors2}
\end{align}
The helicity state are no longer independent but depend on the choice of reference vector:
 \begin{equation}
  u_-(p,m) = \frac{\spA{\fl{p}}{n}}{m} \left(u_+(p,m)\bigg|_{\fl{p} \leftrightarrow n}\right).
  \label{eq:uminus2plus}
\end{equation}
This means that we only need to compute one helicity configuration with general reference vectors for a $t \bar{t}$ pair and we will obtain enough information for all four possible spin configurations.

The helicity amplitudes can be obtained in terms of a set of canonical master integrals as well as in terms of special functions using the framework of finite field arithmetic in \textsc{FINITEFLOW} \cite{Peraro:2016wsq}. The diagram generation, color ordering and spinor-helicity algebra are carried out using QGRAF \cite{NOGUEIRA1993279}, FORM\cite{Kuipers:2012rf}, SPINNEY \cite{Cullen:2010jv} and Mathematica. At this stage, the set of independent topologies is already identified. The numerators are then separated into loop momentum-dependent structures and coefficients dependent on the external kinematics. Then the amplitude is recast into a form suitable for IBP reduction,
and after combining contributions from all numerator topologies we obtain the form of an $L$-loop amplitude in terms of a unique set of Feynman integrals $\tilde{G}$ as follows
\begin{equation}
A^{(L)}(\lbrace p \rbrace)  = \sum_{k} \tilde{c}_{k}(\eps,\lbrace p \rbrace) \; \tilde{G}_{k}(\eps, \lbrace p \rbrace).
\label{eq:calc_ampint2}
\end{equation}
We then combine the bare helicity amplitude with the mass renormalisation counterterms contributions to obtain mass renormalized amplitudes. The integrals that appear in these amplitudes are further reduced to a set of master integrals using IBP identities. The IBP relations are generated in Mathematica with the help of LiteRed \cite{article}. This is solved numerically over finite fields within the FiniteFlow framework using the Laporta algorithm. 
\section{Two-loop NNLO corrections for $t \bar{t}$ pair production}
We start with discussing two-loop NNLO corrections for $t\bar{t}$ production including corrections due to massive fermion loop. The corresponding scattering process involves a pair of top quarks and two gluons 
\begin{align*}
0 \rightarrow \bar{t}(p_1)+t(p_2) +g (p_3)+g(p_4),
\end{align*}
with $p_1^2= p_2^2= m_t^2$ and $p_3^2=p_4^2=0$.
The leading color decomposition of $L$- loop $t\bar{t} gg$ amplitude is 
\begin{align*}
\mathcal{A}(1_{\bar{t}}, 2_t, 3_g, 4_g) = n^L g_s^2 \big[ (T^{a_3} T^{a_4})_{i_2}^{\bar{i_1}} A^{(L)}(1_{\bar{t}}, 2_t, 3_g, 4_g) + (3\leftrightarrow 4) \big]
\end{align*} 
where $n =\frac{m_\epsilon \alpha_s}{4 \pi}$, $\alpha_s =\frac{g_s^2}{4 \pi}$, $m_\epsilon= i (\frac{4 \pi}{m_t^2})^\epsilon e^{-\epsilon \gamma_E}$, $g_s$ is the strong coupling constant and $(T^{a})_j^{\bar{j}}$ are the fundamental generators of $SU(N_c)$.
The partial amplitudes can further be decomposed according to their internal flavour structure:
\begin{align}
A^{(1)}(1_{\bar{t}}, 2_t, 3_g, 4_g) &= N_c A^{(1),1} +N_l A^{{(1)}, N_l} +N_h A^{(1), N_h},\nonumber\\
A^{(2)}(1_{\bar{t}}, 2_t, 3_g, 4_g) &= N_c^2 A^{(2),1} +N_c N_l A^{(2),N_l} +N_c N_h A^{(2), N_h}\nonumber\\ &+N_l^2 A^{(2),N_l^2} +N_l N_h A^{(2),N_l N_h} +N_h^2 A^{(2),N_h^2}.
\label{eq:flamplitude2loop}
\end{align}
Here $N_l$ and $N_h$ are the number of closed light quark loops and heavy quark loops respectively.
\begin{figure}[t]
  \begin{center}
    \includegraphics[width=0.85\textwidth]{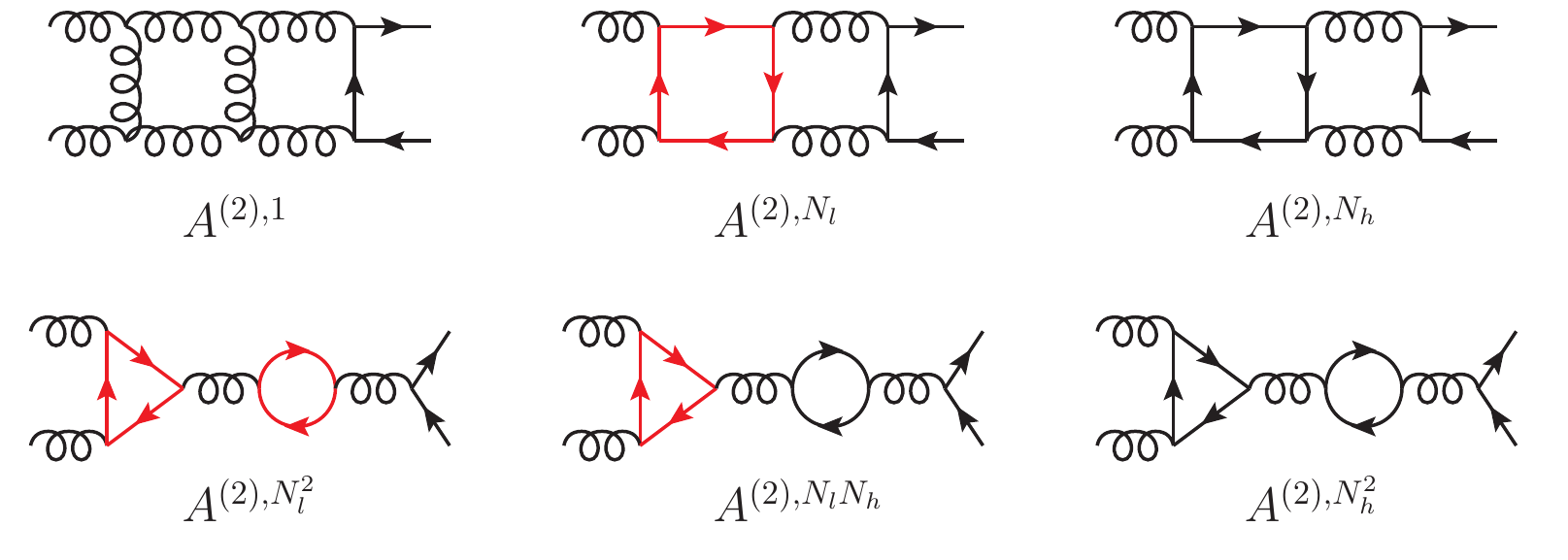} 
  \end{center}
  \caption{Sample Feynman diagrams corresponding to various internal flavour contributions at two loops. 
   Red lines, black spiral lines and black lines represent massless quarks, gluons and top quarks, respectively. This figure is from \cite{Badger:2021owl} licensed under CC-BY 4.0.}
  \label{fig:diagrams2L}
\end{figure}
An $L$-loop $\bar{t}tgg$ amplitude written in terms of the two reference vectors $n_1$ and $n_2$
\begin{align}
  A^{(L)}(1_{\bar{t}}^+,2_t^+,3^{h_3},4^{h_4};n_1,n_2) = m\frac{\Phi^{h_3h_4}}{\spA{\fl{1}}{n_1}\spA{\fl{2}}{n_2}}
  \Bigg(
    &\spA{n_1}{n_2}
    A^{(L),[1]}(1_{\bar{t}}^+,2_t^+,3^{h_3},4^{h_4})\nonumber \\ +
    \frac{\spA{n_1}{3}\spA{n_2}{4}}{\spA34}
    A^{(L),[2]}(1_{\bar{t}}^+,2_t^+,3^{h_3},4^{h_4})\nonumber +
    &\frac{s_{34} \spA{n_1}{3}\spA{n_2}{3}}{\spAA3{14}3}
    A^{(L),[3]}(1_{\bar{t}}^+,2_t^+,3^{h_3},4^{h_4})\nonumber\\ +
    \frac{s_{34} \spA{n_1}{4}\spA{n_2}{4}}{\spAA4{13}4}
    A^{(L),[4]}(1_{\bar{t}}^+,2_t^+,3^{h_3},4^{h_4}&)
  \Bigg),
  \label{eq:ttgg_spinbasis}
\end{align}
where $\Phi$ is a phase factor which depends on the helicities of the gluons. The sub-amplitudes can be computed from 4 different evaluation of the full amplitude with a rational kinematic configuration with 4 choices of the reference vectors. We can use these evaluations to make a linear system which is solved to obtain the sub-amplitudes $A^{(L),[a]}$.
\subsection{Amplitude reduction}
 The master integrals in Eq.~\ref{eq:calc_ampint2} can be expressed as a linear combination of special function monomials $m(f(\lbrace p \rbrace))$ after performing Laurent expansions,
\begin{equation}
A_{\mren}^{(L),h}(\lbrace p \rbrace)  = \sum_{k} \sum_{l=n(L)}^{0} \eps^l \; c^{\mathrm{exp},h}_{kl}(\lbrace p \rbrace) \; m_k\left(f(\lbrace p \rbrace) \right)
+ \cO(\eps),
\label{eq:ampexpanded}
\end{equation}
where $n(L)$ is the power of the deepest pole that can appear in the $L$-loop amplitude. For one- and two-loop cases, $n(1)=-2$ and $n(2)=-4$. 
The finite remainder can be expressed in terms of the monomials as
\begin{align}
F^{(L),h}(\lbrace p \rbrace) & = \sum_{k} \; c^{\mathrm{F},h}_{k}(\lbrace p \rbrace) \; m_k\left(f(\lbrace p \rbrace) \right) + \cO(\eps).
\label{eq:finiteremainder2}
\end{align}
The coefficients $c^{\mathrm{F},h}_{k}(\lbrace p \rbrace)$ are not all independent. We sort all these coefficients by their complexity which is estimated by their total degree determined by a univariate fit \cite{Peraro:2016wsq}. Then we find vanishing linear combinations of these coefficients by solving the linear fit problem for the unknown $y_k$
 \begin{equation}
  \label{eq:linearrels}
  \sum_k y_k\, c^{\mathrm{F},h}_{k}(\lbrace p \rbrace) = 0.
\end{equation}   
Hence we obtain 
\begin{equation}
F^{(L),h}(\lbrace p \rbrace)  = \sum_{k} \; \bar{c}^{\mathrm{F},h}_{k}(\lbrace p \rbrace) \; \bar{m}_k\left(f(\lbrace p \rbrace) \right) + \cO(\eps),
\label{eq:finiteremainderindep}
\end{equation}
where $\tilde{c}_k^{F,h}$ are the independent coefficients of the new monomials $\tilde{m}_k(f)$.

\subsection{Master integrals}
\begin{figure}[t]
  \centering
  \subcaptionbox{The sunrise topology\label{fig:sunrise}}[0.3\textwidth]{\includegraphics[width=2.5cm]{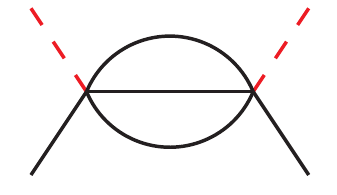}}
  \subcaptionbox{The topbox topology\label{fig:topbox}}[0.3\textwidth]{\includegraphics[width=3cm]{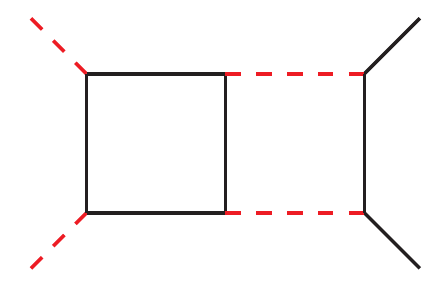}}
  \subcaptionbox{The bubblebox topology\label{fig:bubblebox}}[0.3\textwidth]{\includegraphics[width=2cm]{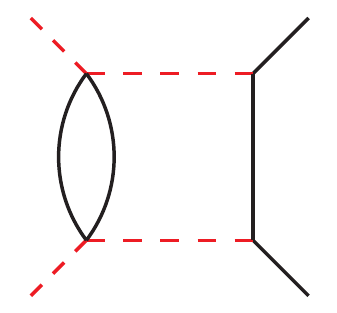}}\\
   \subcaptionbox{The penta-triangle sub-topologies\label{fig:Pentabox}}[1\textwidth]{\includegraphics[width=8cm]{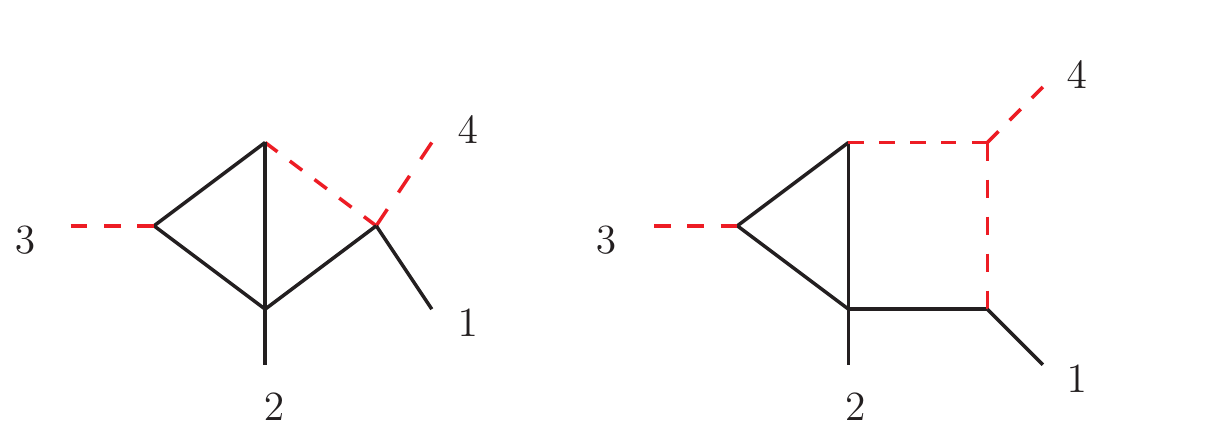}}
\caption{Topologies contributing to $A^{(2),N_h}$ that are associated with the elliptic curves.
Black-solid lines represent massive particles, red-dashed lines represent massless particles. This figure is from \cite{Badger:2021owl} licensed under CC-BY 4.0.} \label{fig:subsectors}
\end{figure}

Now we briefly look at the analytic structure of the contributing master integrals.
The solutions of all one- and two-loop master integrals appearing in the $A^{(2),1}$, $A^{(2),N_l}$, $A^{(2),N_l^2}$, $A^{(2),N_l N_h}$ 
and $A^{(2),N_h^2}$ amplitudes in Eq.~\eqref{eq:flamplitude2loop} are expressible in terms of 
multiple polylogarithms (MPLs) \cite{Mastrolia:2017pfy,Bonciani:2010mn}. On the other hand, for the two-loop amplitude $A^{(2),N_h}$ which involves a single top-quark closed-loop, 
the master integrals are also associated with three elliptic curves \cite{Adams:2018bsn,Adams:2018kez}. 
The elliptic topologies are shown in Fig.~\ref{fig:subsectors}

The results of the first three elliptic topologies come from the topbox family which were already available in \cite{Adams:2018bsn}. For the penta-triangle family, the differential equations for all the 22 master integrals were brought into an eps-factorized form. The results are expressed as iterated integrals with elliptic kernels \cite{Chen:1977oja}.
\subsection{Functional relations}
The functional basis of iterated integrals (the monomials) which are used to express our massive amplitudes has potential redundancies. For example, in order to achieve explicit cancellation of all pole terms, we observed that some functional relations had to be satisfied. These relations look like the following:
\begin{align}
I(a_{3,3}^b, f,...) &=\int a_{3,3}^b\; I(f,...)\;  =\int d \bigg( \psi_1^{(b)} \frac{x (-1+y)}{(-1+x)^2}\bigg)\; I(f,...)\nonumber\\& =\bigg[\psi_1^{(b)}\frac{x(-1+y)}{\pi (-1+x)^2}\; I(f,...) \bigg]_{(0,1)}^{(x,y)} - I\bigg( \psi_1^{(b)}\frac{x (-1+y)}{\pi (-1+x)^2}f,...\bigg).
\label{eq:extrarelations}
\end{align}
The type of relations shown in Eq.~\ref{eq:extrarelations} go beyond shuffle relations. This relation occurs due to the fact that $a_{3,3}^b$ is a total derivative and its primitive vanishes at the lower integration boundary. Such type of relations have also appeared before in literature, for example, in \cite{Remiddi:2017har}. The relations that were used in \cite{Badger:2021owl} originated from the pole cancellation constraint and clearly form a subset of all such  possible relations. Nevertheless, it results in a reduction in the total number of monomials thereby reducing the time taken to reconstruct their coefficients. Hence a faster evaluation of such elliptic amplitudes in the future will require the inclusion of all such types of relations.

\section{Five-point scattering with massive internal propagators}
Now we start discussing the analytic computation of the expansion of the one-loop helicity amplitudes up to $\mathcal{O}(\epsilon^2)$ in the dimensional regulator.  These amplitudes are important ingredients relevant for NNLO computation of $p p \rightarrow t \bar{t}j$. The computation of these amplitudes give us a sense of the complexity that might arise in analytic two-loop $p p \rightarrow t\bar{t} j$ computations.

The two partonic channels for $pp\to t \tb j$ are defined with all
momenta out-going. We write the amplitudes according to the colour decomposition \cite{KLEISS1985235}. Therefore, for the process $0\to
\tb t g g g$ we have: 
\begin{align}
  \mathcal{A}^{(L)}&(1_\tb, 2_t, 3_g, 4_g, 5_g) = g_s^{3+2L} N_\eps^L \bigg\{ \nonumber \sum_{\sigma\in S_3} (t^{a_{\sigma(3)}} t^{a_{\sigma(4)}} t^{a_{\sigma(5)}} )_{i_2}^{\bar{i}_1} A^{(L)}_1(1_\tb, 2_t, \sigma(3)_g, \sigma(4)_g, \sigma(5)_g) \nonumber\\
    + &\sum_{\sigma\in S_3/\mathbbm{Z}_{2}} \delta^{a_{\sigma(3)} a_{\sigma(4)}} (t^{a_{\sigma(5)}} )_{i_2}^{\bar{i}_1} A^{(L)}_2(1_\tb, 2_t, \sigma(3)_g, \sigma(4)_g, \sigma(5)_g) \nonumber\\
    + &\sum_{\sigma\in S_3/\mathbbm{Z}_{3}} \Tr(t^{a_{\sigma(3)}} t^{a_{\sigma(4)}} t^{a_{\sigma(5)}}) \delta_{i_2}^{\bar{i}_1} A^{(L)}_3(1_\tb, 2_t, \sigma(3)_g, \sigma(4)_g, \sigma(5)_g)
    \bigg\},
  \label{eq:colourdecomp_2t3g}
\end{align}
were we use $g_s$ to denote strong coupling and have taken an overall normalisation 
\begin{equation}
  N_\eps = \frac{e^{\eps\gamma_E}\Gamma^2(1-\eps)\Gamma(1+\eps)}{(4\pi)^{2-\eps}\Gamma(1-2\eps)} \,.
  \label{eq:ampnorm}
\end{equation}
The partial amplitudes that appear in the full amplitude as sums over momenta permutations is expressed as $A^{(L)}_i$. The six permutations of 3 gluons is expressed as $S_3$ whereas $S_3/\mathbbm{Z}_{2}$ and $S_3/\mathbbm{Z}_{3}$ are smaller symmetry groups with 3 and 2 elements respectively.
The $SU(N_c)$ colour structures are expressed using the fundamental generators
$(t^a)_i^{\bar{j}}$ with  $a=1,\cdots,8$ being indices of the adjoint representation, while $i,\, \bar{j}=1,2,3$ are indices in the fundamental and anti-fundamental representation respectively.
Similarly, the process $0\to \tb t \qb q g$ is decomposed as
\begin{align}
  \mathcal{A}^{(L)}&(1_\tb, 2_t, 3_q, 4_\qb, 5_g) = g_s^{3+2L} N_\eps^L \bigg\{ \delta^{\bar{i}_4}_{i_1}(t^{a_{5}})^{\bar{i}_2}_{i_3} A^{(L)}_{1}(1_\tb, 2_t, 3_\qb, 4_q, 5_g)\nonumber\\
    + &\delta^{\bar{i}_3}_{i_2}(t^{a_{5}})^{\bar{i}_4}_{i_1} A^{(L)}_{2}(1_\tb, 2_t, 3_\qb, 4_q, 5_g)\nonumber
    - \frac{1}{N_c} \delta^{\bar{i}_2}_{i_1}(t^{a_{5}})^{\bar{i}_4}_{i_3} A^{(L)}_{3}(1_\tb, 2_t, 3_\qb, 4_q, 5_g)\nonumber\\
    - &\frac{1}{N_c} \delta^{\bar{i}_4}_{i_3}(t^{a_{5}})^{\bar{i}_2}_{i_1} A^{(L)}_{4}(1_\tb, 2_t, 3_\qb, 4_q, 5_g)
    \bigg\}.
  \label{eq:colourdecomp_2t2q1g}
\end{align}
Each of the partial amplitudes is further decomposed into a polynomial in $N_c$
and the number of light and heavy flavours, $n_f$ and $n_h = 1$ respectively.
We reconstruct analytically in the minimal set of six
on-shell variables while making use of the following basis for spin structures,
\begin{align} \label{eq:spindecomp}
  A_x^{(L)}(1_t^+,2_\tb^+,3^{h_3}&,4^{h_4},5^{h_5};n_1,n_2) =
  m_t \Phi(3^{h_3},4^{h_4},5^{h_5})\nonumber\\&
  \sum_{i=1}^4 \Theta_i(1,2;n_1,n_2) A_x^{(L),[i]}(1_t^+,2_\tb^+,3^{h_3},4^{h_4},5^{h_5}).
\end{align}
The phase factor $\Phi$ accounts for the massless
parton helicities, four basis functions $\Theta_i$ contain the spin dependence of
the top-quark pair and the associated subamplitudes $A_x^{(L),[i]}$. The dependence on the arbitrary reference vectors used to define positive massive fermions is contained in functions $\Theta$. 
\subsection{Amplitude reduction}
After identifying the integral topologies from the loop momentum dependent propagators, as explained in section \ref{sec:method}, the coefficients of these numerators are computed using the momentum twistors parametrisation. We compute the diagram numerators using a symbolic value for the spin dimension $d_s = g_\mu^\mu$ and the amplitude is presented as
\begin{align}
  A_x^{(L),[i]} = A_x^{(L,0),[i]} + (d_s-2) A_x^{(L,1),[i]}.
\end{align}
After this, the diagram numerators are reduced to master integrals using IBP implemented in \textsc{FINITEFLOW}. We also perform wave-function renormalisation to obtain a gauge invariant result.
The coefficients of the master integrals depend on the dimensional regularization parameter $\epsilon$ and six free parameters in the rational phase-space. The polynomial degree increases after rationalizing the phase-space therefore we apply linear relations and a univariate partial fractioning to the master integrals coefficients before reconstruction \cite{Badger:2021imn}. 

\subsection{Master integrals}
\begin{figure}[h]
  \begin{center}
    \includegraphics[width=5.5cm]{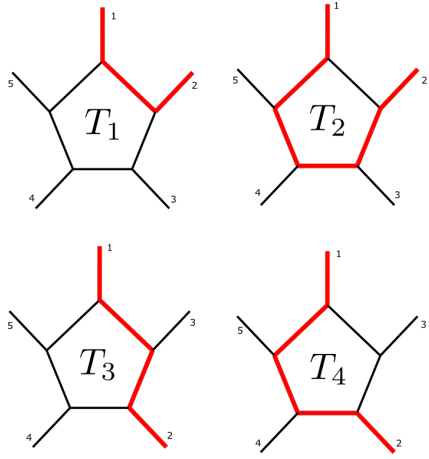}
  \end{center}
  \caption{The four distinct one-loop integral topologies appearing in the $pp\to t\tb j$ amplitudes. Black lines denote massless particles while red lines denote massive particles. This figure is from \cite{Badger:2022mrb} licensed under CC-BY 4.0.}
  \label{fig:pentagontopologies}
\end{figure}

The amplitudes contain four distinct pentagon topologies shown in Fig.~\ref{fig:pentagontopologies}. 
Across these topologies there are 130 master integrals. We compute these master integrals using differential equations after obtaining a canonical form. The square roots that appear in their systems of differential equations are shown below:
\begin{align}
  \beta(a_1,m^2) &= \sqrt{1-\frac{4 m^2}{a_1}},\qquad
  \Delta_3\left(P,Q\right) \;= \sqrt{(P\cdot Q)^2 - P^2 Q^2}, \nonumber \\
  \operatorname{tr}_5 &= \operatorname{tr}_5(3,4,5,1) = \sqrt{\operatorname{det}G(p_3,p_4,p_5,p_1)}, 
  \label{eq:sqrt}
\end{align}
where $a_1$ is a function of the kinematic invariants, $P$ and $Q$ are momenta and
$G_{ij}(\vec{v}) = 2 v_i\cdot v_j $ is the Gram determinant matrix. To compute the differential equations we employ the technique of generalized power series expansion  \cite{Moriello:2019yhu} to solve them numerically. This technique would be useful also for two-loop integrals even in the presence of non-polylogarithmic forms. We solve the system of differential equations for the 130 master integrals using the implementation of this technique in DiffExp \cite{Hidding:2020ytt}.  We also obtain the analytic boundary conditions for all the master integrals but for the pentagon topology in $T_2$ for which we obtain expressions in terms of one-parameter integrals.
\section{Results}
For the 2-loop amplitudes for $pp \rightarrow t\bar{t}$, we obtained analytic amplitude level expressions using the massive spinor-helicity formalism for the first time in \cite{Badger:2021owl}. The algebra is compatible to be performed with finite fields arithmetic. The integrals contain elliptic functions and result in a complicated analytic structure \cite{Chaubey:2021ret}. The results of the elliptic master integrals are expressed in terms of iterated integrals over kernels associated with three elliptic curves. These monomials containing iterated integrals obey additional functional relations. Understanding these relations in more detail is needed in the future to reduce the complexity of final expressions in the amplitudes. For the  1-loop $t\bar{t}j$ case, the analytic form for the partial helicity amplitudes has been explicitly obtained up to $\mathcal{O}(\epsilon^2)$ after breaking them into sub-amplitudes in \cite{Badger:2022mrb}. We have collected the coefficients appearing in the subamplitudes and applied linear relations between them. These subamplitudes are expressed in terms of the master integrals which are evaluated  using generalized series expansion after obtaining a canonical form. Both of the works presented here motivate further studies into analytic as well as semi-analytic approaches to high-precision $pp\rightarrow t \bar{t}$ and $pp \rightarrow t \bar{t}j$ amplitudes and cross sections.

\section{Acknowledements}
EC would like to thank her collaborators Simon Badger, Matteo Becchetti, Heribertus Bayu Hartanto, Robin Marzucca and Francesco Sarandrea for useful comments and discussions.
EC receives funding from the European Union's Horizon 2020 research and innovation programmes \textit{High precision multi-jet dynamics at the LHC} (consolidator grant agreement No. 772009). 
\bibliographystyle{JHEP}
\bibliography{ll.bib}
\end{document}